# Structure of the Solar Dust Corona and its Interaction with the other Coronal Components

Y.Y. Shopov[a*], D. A. Stoykova[a], K. Stoitchkova[a], L.T.Tsankov[a], A. Tanev[a], Kl. Burin[a], St. Belchev[b], V. Rusanov[a], D. Ivanov[a], A.Stoev[c], P. Muglova[d], I. Iliev[d]

[a]*Faculty of Physics, University of Sofia, J. Bourchier 5, Sofia 1164, Bulgaria*
[b]*Faculty of Chemistry, University of Sofia, J. Bourchier 1, Sofia 1164, Bulgaria*
[c]*People Astronomical Observatory "Yu. Gagarin", Stara Zagora, Bulgaria*
[d]*Solar- Terrestrial Influences laboratory of Bulgarian Academy of Sciences, Sofia 1000, Bulgaria*

**Abstract.**

We developed a new technique for registration of the far solar corona from ground based observations at distances comparable to those obtained from space coronagraphs. It makes possible visualization of fine details of studied objects invisible by naked eye. Here we demonstrate that streamers of the electron corona sometimes punch the dust corona and that the shape of the dust corona may vary with time.

We obtained several experimental evidences that the far coronal streamers (observed directly only from the space or stratosphere) emit only in discrete regions of the visible spectrum like resonance fluorescence of molecules and ions in comets. We found that interaction of the coronal streamers with the dust corona can produce molecules and radicals, which are known to cause the resonance fluorescence in comets.

## 1. Introduction

From most of the locations on the Earth **white light corona** can be observed only during total solar eclipses because its intensity is much lower than the brightness of the sky. Edges of the corona gradually disappear in the background of the sky. So the size of the corona greatly depends on the spectral region of observations and clearness of the sky. It has several components emitting in the entire visible region of the spectrum:

*K- (Electron or continuum) corona* is due to scattering of sunlight on free high-energy electrons. Its spectrum is continuous. It dominates the corona brightness at distances closer than 2.3 solar radii (see Fig. 1) and it has variable structures depending on the level of solar activity. It has a distinct 11 years cycle.

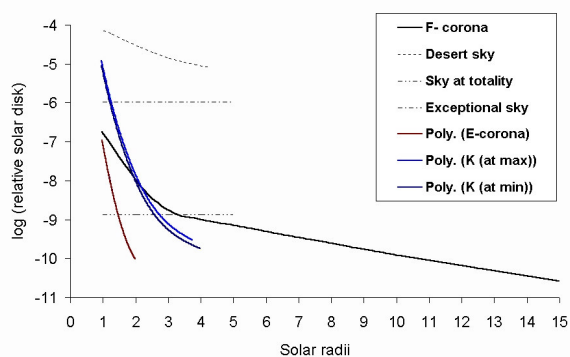

Fig.1 Decreasing of the intensity of the components of solar corona with the distance from the Sun

*F- (Fraunhofer or Dust) corona* is due to scattering of sunlight on dust particles. It has Fraunhofer spectra with absorption lines. F-corona usually has oval shape (Figs. 2, 3). Its intensity falls down very slowly with the distance

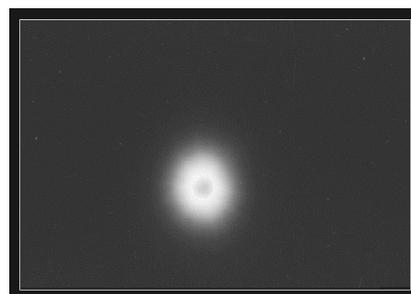

Fig.2 F- corona during the solar eclipse in 1999 in near Infrared light (690-950 nm)

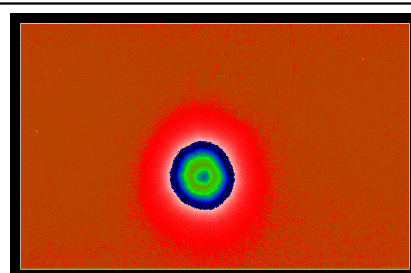

Fig.3 Color map of the density of F- corona from fig.2 during the solar eclipse on 11.08.1999. The green followed by blue color represent higher density

from the Sun and it predominates over K-corona at distances higher than 3 solar radii (Fig. 1). Figure 1 represents the data of Golub and Pasachoff (1997) for K- and E- corona and our data for F-corona up to 15 solar radii far from the Sun. Prasad (1995) suggested that the dust ring varies with the solar cycle but Ohgaito et al. (2002) claim the opposite. Ragot and Kahler (2003) demonstrated that the dust corona interacts with both the particles and fields of the coronal mass ejections (CMEs) and the solar wind.

*Thermal (T) corona* is produced by thermal emission of dust particles heated by the Sun. The spectrum of emission of the T-corona is equal to the spectrum of heated black body. Its maximum is in the near infrared (NIR) region of the spectrum (Fig. 4).

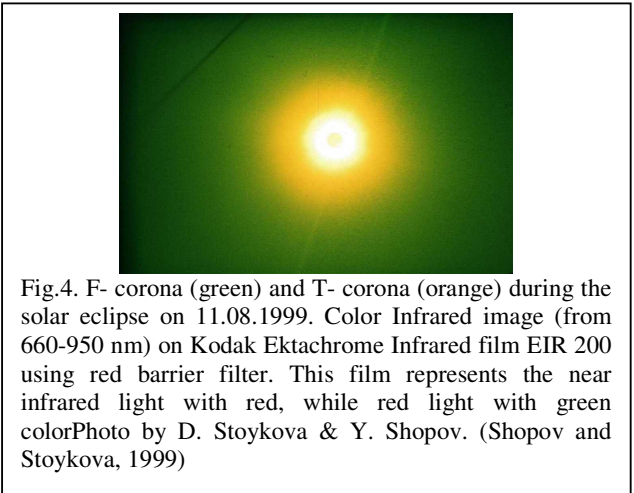

Fig.4. F- corona (green) and T- corona (orange) during the solar eclipse on 11.08.1999. Color Infrared image (from 660-950 nm) on Kodak Ektachrome Infrared film EIR 200 using red barrier filter. This film represents the near infrared light with red, while red light with green colorPhoto by D. Stoykova & Y. Shopov. (Shopov and Stoykova, 1999)

The shape of the T-corona (fig.5, right) is a little bit different than that of the F-corona (fig.5, middle, fig. 6).

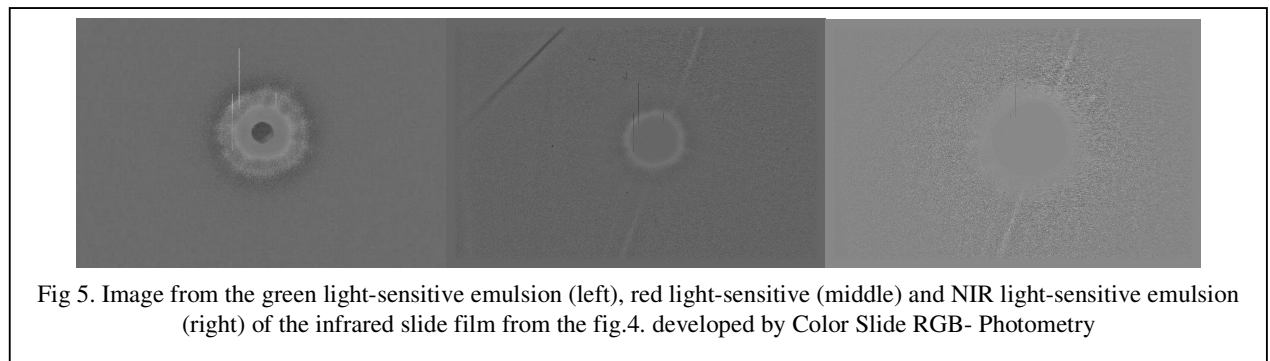

Fig 5. Image from the green light-sensitive emulsion (left), red light-sensitive (middle) and NIR light-sensitive emulsion (right) of the infrared slide film from the fig.4. developed by Color Slide RGB- Photometry

Solar corona has also several components emitting line emission spectrum:

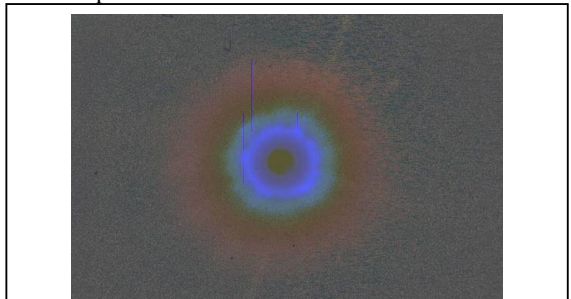

Fig 6. Sum of the images from the 3 color layers of slide film from the fig.5.K- corona is represented

*E- (Emission) corona* is due to a line emission spectrum of highly ionized atoms of Fe, Ni and Ca (Golub and Passachoff, 1997), (Figs. 7,8). Its intensity falls down very rapidly with the distance far from the Sun (Fig. 1) and it is visible up to 2 solar radii in monochromatic light.

These coronal components were established long ago.

The aim of this work is to study the relations and interactions between the different components of the solar corona

**2. Experimental Techniques:**
**The New Technique of "Color Slide RGB-Photometry"**

The new technique of "Color Slide RGB-Photometry" recently suggested by (Shopov et al., 2002b) consists of extraction of separate images from each of the 3 color layers of a slide film. Each of these images is obtained from different emulsion layer with different spectral sensitivity. Then they are filtered separately. Filtration procedure consists of the following:

1. Each of the images is smoothed by median filtration with a window of 25- 1200 pixels depending on the size of the image (in pixels) and the scale of the searched details of the image.
2. Smoothed image is subtracted from the original image
3. Residual image is rescaled in 256 levels of the gray density in order to visualize all details in the obtained image

The obtained 3 images from the 3 emulsion layers of the film visualize much more details (Fig. 5), invisible in the initial photograph (Fig. 4). They can be summed to produce a new filtered color image (Fig. 6). The final image exhibits more fine details of the recorded object than the initial image (Fig. 4) both in color and structure.

This new technique allows research and estimation of the differences in the color and structure of elements of the images in different spectral regions of light. It allows also visualization of fine details invisible for naked eye (due to undetectably low differences of their brightness from the neighbor parts) of the studied objects. Human eye can distinguish only about 256 levels of the gray density, while the photo emulsion can record unlimited number of levels of the gray density, which can be measured in 65536 ($2^{16}$) or more levels of the gray density by a scanner or scanning microdensitometer. Color Slide RGB-Photometry technique allows visualization of the information recorded in such images.

**3. Study of the shape of the Dust Corona**

Traditionally it is believed that F-corona has shape of a slightly extended ellipsoid around the Sun and that it does not vary with time. In fact, dust particles evaporate close to the Sun producing a large cavity in the dust distribution. So, practically the dust corona has the shape of a hollow ellipsoid with the Sun in the center (Ohgaito et al., 2002).

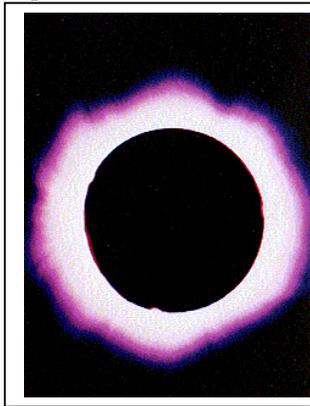

Fig.7 NIR emission of Fe XI during the solar eclipse in 1999 represented in red by Color Kodak Ektachrome Infrared film EIR 200 (image registered from 390-950 nm), after Shopov et al. (1999a). Blue color is due to UV emission.

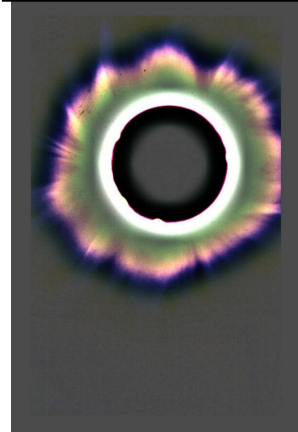

Fig.8 E- corona during the solar eclipse in 1999 Color Infrared image processed by Slide RGB- Photometry. Red emission in Hα is represented with green color, but NIR emission of Fe XI with red color by the EIR 200 film

Our observations confirm this hypothesis. The dust corona is best represented in the image on the red-light sensitive emulsion of the infrared slide film developed by Color Slide RGB- photometry (fig 5, middle). This image demonstrates also that the streamers of the electron corona punch the dust corona. Streamers of the K-corona, which are visible on (fig 5, left) produce cavities in the distribution of the dust (fig 5, middle). These streamers of heated and more dense plasma evaporate or scatter the dust more efficiently that the rest of the solar corona. So penetrating deeper in the dust cloud they produce uneven internal surface of the dust corona which is visible on fig. 5, 6. Following observations of the total solar eclipse on the 29 March 2006 (fig.9) approved definitely, that streamers of the electron corona sometimes punch the dust corona and that the shape of the dust corona also varies from eclipse to eclipse as the other coronal components. Before it was considered to be constant with time (i.e. Bakulin et al., 1977).

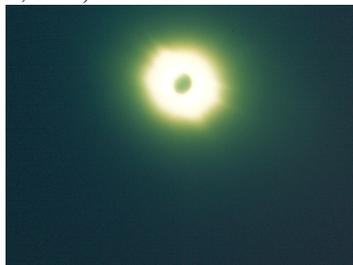

Fig.9. F-(green) and T-corona (orange) at the solar eclipse in 2006
Photo by D. Stoykova & Y. Shopov

Fig.5 (left) suggests that the cavity in the dust corona accommodate the K- corona. This cavity is represented on Fig.5 (left) by the entire region inside the dark ring. The K- corona (which shape is well known from other observations of the solar eclipse on 11 Aug. 1999) is visible inside the cavity in the dust corona on Fig.5 (left) and fig. 6. The next visible structure in the same figures is an intermediate region in which streamers of the K-corona penetrate the dust surrounding them. The most outer region is the dust corona. The cavity on Fig.5 (middle) represents the dust- free region around the Sun. These observations suggest that in fact the internal surface of the dust corona depend on the shape of the electron corona. It has forms negative to the shape of the streamers of the K-corona.

**4. New Coronal Components Produced by Interactions of the Dust Corona with the Streamers of the K-Corona**

*S- (Sublimation) corona* was found in the last years by Gulyaev (2002) but it is still not known widely. It consists of emission of low ionized atoms (like Ca-II) produced by sublimation of dust particles in relatively cold parts of the corona (Fig. 10).

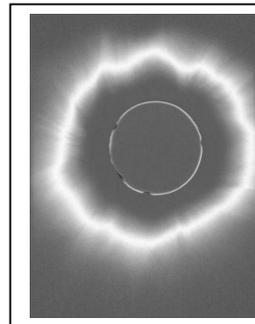

Fig.10. Bright CaII emission of the S- corona, 11.08.1999 after Shopov et al. (2002b)
Photo by A. Tanev & Y. Shopov

The coronal streamers of the K- corona (Fig. 5) punch the internal surface of the dust corona because they evaporate the interplanetary dust. This interaction between the dust and the heated plasma of the electron corona produces material for formation of other coronal components. It produces CaII ions, which build the S-corona. About 50% of the coronal dust consist of $CaCO_3$. Its thermal decomposition by hot plasma of the streamers of the K-corona follows the reaction:

$CaCO_3 \rightarrow CaO + CO_2$ (1)

Further heating produce evaporation or sublimation of CaO and supplies CaII. Its emission in 393.4 nm has Doppler shift (Gulyaev, 2002) suggesting moving of ions with speed similar to this of the solar wind and acceleration from 420 km/s (at 5) to 870 km/s (at 10 solar radii). Such acceleration with the distance from the Sun is typical for the solar wind (Vaisberg, 1986). These facts suggest that CaII ions are moved by the solar wind rather than by the light pressure, which drops with the distance from the Sun.

K-, E-, F-, T- and S- coronal components are visible together in the corona from the ground during total eclipses (Shopov, 2003).

The most mysterious component of the corona is the **"giant coronal streamers"** observed only from the LASCO coronagraph and from stratospheric flights during total eclipses (Martinez, 1978, Keller et al., 1981, Mulkin, 1981, Shopov et al., 1999b, 2003). Its shape and properties are different than these of any other component of the corona. During the last years arose many evidences demonstrating that its nature is the same as that of plasma tails of the comets- Fluorescence of ionized gas molecules due to their interaction with the sunlight and solar wind.

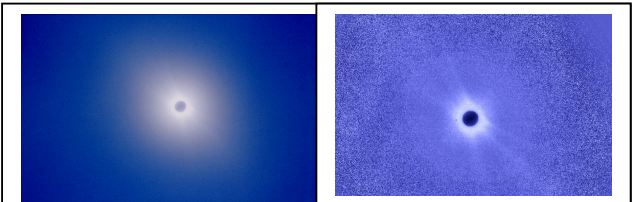

Fig 11. (Left) Total Solar Eclipse on the 29 March 2006 registered on Agfachrome Precisa 200 ISO slide film using Sonnar 2.8/ 180 mm lenses and 1 second exposure. Photo by Katerina Stoichkova and Yavor Shopov.
(Right) Image obtained from the red emulsion layer of the slide film of the left image using "Color Slide RGB-Photometry" technique (Shopov et al., 2002b). Blue color of the image is artificially introduced.

During the totality, the intensity of the far solar corona becomes lower than the intensity of the sky glow at large distances, far from the Sun. Therefore far solar corona disappears gradually in the dark sky [Fig. 11 (left)]. Even during the totality the sky is blue. Depending on the sky transparency the solar corona can be seen up to 4–5 solar radii by eye and even less on film because it is less sensitive in the red end of the spectra. The total solar eclipse on the 29 March 2006 did not make exception. Simultaneous observations from the space on LASCO-C3 coronagraph registered the solar corona up to 30 solar radii (fig.12, right).

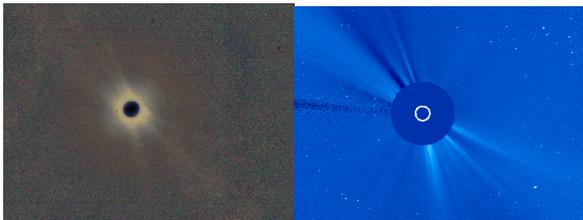

Fig. 12. Sum of the images from the 3 color layers of the slide film from the upper image developed by Color Slide RGB-Photometry (Shopov et al., 2002b) (left), compared to the image from the Large Angle Space Coronagraph (LASCO)- C3 coronagraph registered simultaneously from the space at 10.51 UT on the 29 March 2006. Both images represent the same portion of the sky and solar corona. LASCO image has been rotated to have the same orientation as the ground image

The reasons why it can not be observed directly from the Earth's surface are:
(*a*) Sky glow during the totality,
(*b*) Atmospheric halo,
(*c*) Optical halo of the telescope/tele-lenses produced by much brighter internal corona.
The last two interferences can be removed by using of the new technique of Color Slide RGB-Photometry.
The giant coronal streamers visible on Fig. 11 (right) are registered only in the red and green sensitive emulsion layers but are not registered in the blue light sensitive emulsion layer of the slide film (Fig. 13). It suggests that the giant coronal streamers were not emitting in the blue region during this eclipse. This fact is rather surprising because during the eclipse on the 11 August 1999 they were registered only in the blue and green light sensitive emulsion layers. They were not registered in the red-light sensitive emulsion layer of the same type of slide film (Shopov et al., 2002). This fact suggests that the giant coronal streamers emit only in certain spectral regions of the visible light and these regions change with time. This phenomenon can not be detected by the SOHO satellite, because it does not have any visible and NIR- light spectrograph but LASCO coronagraph registers only integral light images in the spectral region of 500–1000 nm (LASCO C3 Quantum Efficiency,
http://lasco-www.nrl.navy.mil/content/tech/QE/c3_qe.pdf).

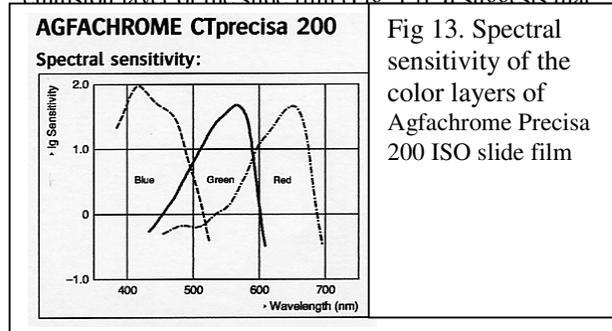

Fig 13. Spectral sensitivity of the color layers of Agfachrome Precisa 200 ISO slide film

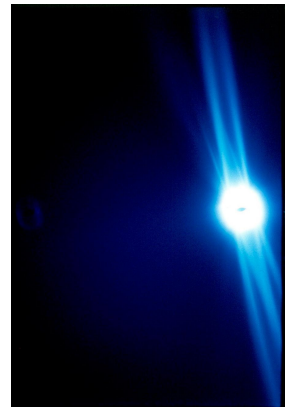

Fig 14. Giant coronal streamers registered during the solar eclipse in 1999 from a sky fighter Mig- 21 on Kodak Ektachrome IR film EIR 200. Blue color is due to blue or green emission

Additional evidences that giant coronal streamers emitted only blue and green light during the eclipse on the 11 August 1999 (Fig. 14) were obtained directly from a hypersonic fighter flying at 13 000 m altitude with speed of 1600–1700 km/h (Shopov et al., 1999b). There were two equivalent optical systems mounted parallely on the same mount in the airplane. One of them was working in the regions of 500–890 nm and 400–890 nm while the other only in the region of 630–930 nm. These streamers emitted only light at wavelengths shorter than 630 nm. We can make this affirmation because the second camera on the airplane (working in the region of 630– 930 nm) did not register them. The photo of the solar corona presented on Fig. 14 and other similar images [see Astronomy 28 (2000) 134] were recorded (at distance over 25 solar radii far from the Sun) on infrared specto-zonal film Kodak EIR 200, with a yellow filter (Kodak No. 12). This emulsion represents green and blue light with such deep blue color, which means that original light emission of the far parts of the coronal streamers was green and blue. Otherwise white light should be represented with white color in the presence of the used yellow filter Kodak No. 12 (standard for this type of film).

There were very few similar observations from stratospheric flights during total eclipses (Martinez, 1978, Keller et al., 1981, Mulkin, 1981). They registered the giant coronal streamers on black and white films only, so they did not obtain any instrumental record of the color of the streamers. Therefore they did not mention the observed phenomenon.

Another process of fluorescence emission in the solar corona has been observed by Habbal et al. (2003). They observed **IR-fluorescence** of silicon nanoparticles at 1070–1080 nm in coronal holes at distances $> 1R$. This silicon can be also produced by thermal decomposition of $SiO_2$ particles of the dust corona.

## 5. Possible Origin of the Far Coronal Streamers

Obtained data suggest that far coronal streamers do not belong to the white light corona. They probably emit only in broad- band lines of resonance fluorescence of molecules and ions. These molecules can be carried far from the Sun by the solar wind or coronal mass ejections like the ions of the S-corona. Our observations of the total solar eclipse on the 29 March 2006 support such hypothesis. $CO_2$ produced by reaction (1) can produce various molecules and ions known to have resonance fluorescence in comets, like $CO^+$, $C_3$ and $C_2$ (fig. 15).

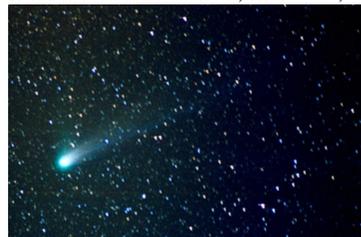 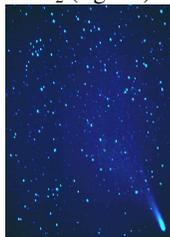

Fig. 15. Comet C2001Q4 (NEAT) on 23 May 2004 Green color of the coma is due to fluorescence of $C_2$ molecule.( Photo by M. Tomov, A. Tanev and Y. Shopov

Fig. 16. Color Infrared image of Comet Ikeya- Zhang on Kodak Ektachrome Infrared film EIR 200 with Kodak-12 yellow filter, obtained with the same instrument, film and

Evidently such molecules should generate even greater emission close to the Sun due to the higher intensity of the sunlight which excites their fluorescent emission. The similar spectral composition (fig.14, 16) suggests similar origin of the emission of comets and of the far coronal streamers.

Resonance fluorescence of molecules should visualize the leakage of the solar wind out of the Sun because solar wind carries the sources of these molecules. In fact animations of the development of the giant coronal streamers with time looks like visualization of gusts of solar wind. NASA found strong correlation between the intensity of coronal mass ejections (like CME on the 26 of December 2001) and the density of the high-energy particles of the solar wind (http://lheawww.gsfc.nasa.gov/~reames/gsfc3.html, http://lheawww.gsfc.nasa.gov/~reames/DARK7.HTML).

Based on these considerations we suggest a working name for this part of the corona– **"fluorescent corona" or "Fl-corona"**.

## 6. Comparison of the properties of the coronal components

Comparison of the properties and behavior of the coronal components is given in table 1.

## 7. Conclusions

We demonstrated that the streamers of the electron corona sometimes punch the dust corona producing uneven internal surface of the dust corona. We also demonstrated that the shape of the dust corona might vary from eclipse to eclipse but it needs more work to clarify if it is relevant to the solar cycle or not.

Far coronal streamers observed from the orbital coronagraph LASCO and from stratospheric flights during total eclipses do not belong to the white light corona. They emit only in discrete regions of the visible spectra like resonance fluorescence of molecules and radicals due to their interaction with the solar wind and sunlight. Such molecules and radicals can be originated by evaporation of the dust corona.

Dust corona supplies $CO_2$, which is producing the fluorescent corona. It also supplies CaO, producing the sublimation corona by thermal decomposition of $CaCO_3$ due to its heating by the hot plasma of the streamers of the electron corona. Probably it also supplies silicon from the $SiO_2$ in the dust by the same mechanism producing the IR-fluorescence corona.

### Acknowledgements

Authors express great thanks to Mr. Dimitar Petrov and Mr. Mikhail Tomov for the priceless help in registration of the images and to Dr. Miltcho Tsvetkov for the microdensitometry and preparation of the color map of the F-corona.

| **Coronal components** | **Origin** | **Region of existence** | **Emission spectra** | **Shape** | **Cycle** | **Interactions** |
|---|---|---|---|---|---|---|
| *K-(electron) corona* | photospheric light scattered by the free electrons | follows the magnetic lines configuration | continuous white sunlight spectra | depends on the magnetic field | strong solar activity cycle | evaporates dust of the F-corona to produce S-, Fl - and IRF- corona |
| *F-(dust) corona* | reflecting of sunlight on dust particles | outer corona > 2.5 R up to the Earth. Best visible in IR | Fraunhofer white sunlight spectra | hollow sphere punched by the streamers of K-corona | No | sublimates or evaporates by streamers of the K-corona |
| *T-(thermal) corona* | thermal emission of dust heated by the sunlight | outer corona > 2.5 R | continuous thermal emission, mainly in NIR | Variable from eclipse to eclipse. Best visible in NIR | | K-corona evaporates dust of the T-corona |
| *E-(emission) corona* | emission of highly-ionized ions | inner corona | linear emission of highly-ionized ions | depends on the distribution of highly- ionized ions | | |
| *S-(sublimation) corona* | sublimation of dust of the F-corona | > 0.1- 0.2 R up to 10 R | linear emission of low-ionized ions at 170-870 km/h | negative streamers of the K-corona | follows cycles of K-corona? | K-corona evaporates dust of the F-corona to produce S-corona |
| *Fl- corona (fluorescence) corona* | Resonance fluorescence of molecules and free radicals | giant coronal streamers following solar wind or CME | broad emission lines of resonance fluorescence | giant coronal streamers up to 30R. Highly and rapidly variable | follows streams of the solar wind | It is produced by interactions of the K- and F-corona, solar wind and sunlight |
| *IRF- (Infrared fluorescence) corona* | Infrared fluorescence of silicon nanoparticles | coronal holes at distances >1R | narrow IR-fluorescence lines | follow coronal holes | ? | It is produced by interaction of the K- and F-corona and sunlight |

Table 1. Comparison of the properties of the coronal components